\documentclass[12pt]{iopart}

\usepackage{iopams,graphicx}  
\begin{document}

\title{The pressure-induced enhancement of superconducting properties of single-crystalline FeTe$_{0.5}$Se$_{0.5}$}

   \author{J Pietosa$^1$, D J Gawryluk$^1$, R Puzniak$^1$, A Wisniewski$^1$, J Fink-Finowicki$^1$, M Kozlowski$^{1,2}$ and M Berkowski$^1$}
   \address{$^1$ Institute of Physics, Polish Academy of Sciences, Aleja Lotnikow 32/46, PL-02-668 Warsaw, Poland}
   \address{$^2$ Tele and Radio Research Institute, Ratuszowa 11, PL-03-450 Warsaw, Poland}
   
\ead{pietosa@ifpan.edu.pl}

\begin{abstract}
The pressure dependence, up to 11.3 kbar, of basic parameters of the superconducting state, such as the critical temperature (${T}_{\rm c}$), the lower and the upper critical fields, the coherence length, the penetration depth, and their anisotropy, was determined from magnetic measurements performed for two single-crystalline samples of FeTe$_{0.5}$Se$_{0.5}$. We have found pressure-induced enhancement of all of the superconducting state properties, which entails a growth of the density of superconducting carriers. However, we noticed more pronounced increase in superconducting carrier density under pressure than that in the critical temperature what may indicate an appearance of a mechanism limiting the increase of ${T}_{\rm c}$ with pressure. We have observed that the critical current density increases under pressure by at least one order of magnitude.
\end{abstract}

\maketitle

\section{Introduction}

The discovery of superconductivity in the Fe-based oxypnictide compounds \cite{1} has sparked tremendous interest and opened up new perspectives in the field of superconductivity \cite{2,3,4,5,6}. Until now, the following groups of Fe-based superconductors are known: REOFeAs, ("1111", RE = rare earth) \cite{1},  AFe$_{2}$As$_{2}$ (''122'', A = alkaline earth) \cite{2}, LiFeAs (''111'') \cite{3}, Fe(Se,Ch) (''11'', Ch = S, Te) \cite{4,5}, and Sr$_{2}$MO$_{3}$FePn (''21311'', M = Sc, V, Cr, and Pn = pnictogen) \cite{6}. Within the ''11'' group, pure FeSe exhibits superconductivity below ${T}_{\rm c}$ $\approx$ 8 K \cite{4}. The tetragonal compounds FeSe and FeTe$_{1-x}$Se$_{x}$ have a quite simple structure, with Fe and Te/Se layers additionally with Fe excess, alternating along the \textit{c}-axis \cite{4,7,8}. These compounds have attracted much interest because of their similarities to the high-${T}_{\rm c}$ iron pnictides. The values of critical temperature in these compounds are much lower than those of FeAs-based superconductors. However, the simplicity of structure and similarity in the Fermi surface among pnictides make studies of FeTe$_{1-x}$Se$_{x}$ potentially useful for understanding of the mechanism of superconductivity in all Fe-based oxypnictides. Partial substitution of Te for Se leads to an increase of ${T}_{\rm c}$ up to about 14 K for Fe$_{1-y}$Te$_{1-x}$Se$_{x}$ with $0.4 < x < 0.8$ and $y\approx0$ \cite{7,9}.

The application of external pressure (\textit{P}) to the pure FeSe has led to a raise of ${T}_{\rm c}$ \cite{10,11,12,13} up to 36.7 K at 89 kbar \cite{11}. Interestingly, a similar high ${T}_{\rm c}$ $\approx$ 30 K is attained in the iron-selenide family A$_{x}$Fe$_{2-y}$Se$_{2}$ by intercalating alkaline earth atoms (A = K, Rb, Cs) between the FeSe layers \cite{14,15,16}. However, ${T}_{\rm c}$ is found to decrease with pressure and is fully suppressed at 90 kbar for K$_{x}$Fe$_{2-y}$Se$_{2}$ \cite{17} and at 80 kbar for Cs$_{x}$Fe$_{2-y}$Se$_{2}$ \cite{18}. The critical temperature is very weakly dependent on pressure below 10 kbar, suggesting that ${T}_{\rm c}$ is almost independent of small variations of the lattice constants.

In the case of FeTe$_{0.5}$Se$_{0.5}$, the ${T}_{\rm c}$ increases with \textit{P} \cite{19,20,21} up to 26.2 K for \textit{P} = 20 kbar \cite{19}. It is interesting that at higher pressures (above 20 kbar), ${T}_{\rm c}$ decreases \cite{19,20} down to zero at about 110 kbar \cite{20}. This was explained as a result of the pressure-induced disordering of the Fe(Se,Te)$_{4}$ tetrahedra, noticed at 110 kbar in X-ray diffraction studies at room temperature \cite{20,22}.

However, the pressure studies of superconductivity in Fe(Se,Ch) system were limited mainly to a tuning of ${T}_{\rm c}$. The pressure dependence of the upper critical field (${H}_{\rm c2}$) was investigated for polycrystalline FeSe, only \cite{10}. Still, nothing is known about the pressure dependence of the lower (${H}_{\rm c1}$) and the upper critical fields, and their anisotropies for single-crystalline FeTe$_{1-x}$Se$_{x}$. There is a lack of data on the pressure impact on the critical current density (${j}_{\rm c}$) in FeTe$_{1-x}$Se$_{x}$. Since we have established earlier \cite{23} that the sharpness of a transition to the superconducting state in FeTe$_{1-x}$Se$_{x}$ is evidently inversely correlated with crystallographic quality of the crystals, we decided to perform pressure studies of two FeTe$_{0.5}$Se$_{0.5}$ single crystals of significantly different crystallographic quality.

The lower critical field, related to London penetration depth, provides information about the density of superconducting carriers. The upper critical field, directly related to the coherence length, and its temperature dependence, provide some information about pairing mechanism and pairing strength. Both microscopic quantities, together with the critical current density, are important for application purposes as well. In this paper, the pressure dependence of the lower and the upper critical fields and of the critical current density in FeTe$_{0.5}$Se$_{0.5}$, is presented. The hydrostatic external pressure, up to 11.3 kbar, has led to a more pronounced increase in superconducting carrier density than that in the critical temperature, what may indicate an appearance of a mechanism limiting the increase of ${T}_{\rm c}$ with pressure.

\section{Synthesis and experimental techniques}

Single crystals of nominal composition FeTe$_{0.5}$Se$_{0.5}$ have been grown using Bridgman's method. The studied samples were prepared from stoichiometric quantities of iron chips (3N5), tellurium powder (4N), and selenium powder (pure). All of the materials were weighed, mixed and stored in a glove box in argon atmosphere. Double walled evacuated and sealed quartz ampoule with starting materials was placed in a furnace with a vertical gradient of temperature equal to $\sim$1.2 $^{\rm o}$C/mm for the Sample I and $\sim$0.6 $^{\rm o}$C/mm for the Sample II. The material was synthesized for 3 h at temperature 730 $^{\rm o}$C and next temperature was risen up to 920 $^{\rm o}$C. After melting, the temperature was held for 3 h, then the sample was cooled down to 500 $^{\rm o}$C with a rate of 1.5 $^{\rm o}$C/h (Sample I) or 3 $^{\rm o}$C/h (Sample II) and next to 200 $^{\rm o}$C with a rate of 60 $^{\rm o}$C/h for both samples, and finally cooled down with the furnace to room temperature. As a result, we have obtained two single crystals with different crystallographic quality. In our case, the crystal quality was determined by the $\Delta\omega$ value, describing the full width at half maximum (FWHM) of the 004 X-ray diffraction peak, obtained in the $\omega$ scan measurements, since changes in the \textit{c}-axis lattice constant are very sensitive to the variation in chemical composition of studied materials \cite{23}. The 004 peak is relatively intense and appears at sufficiently large angles to get a good angular resolution. The crystals, with $\Delta\omega$ values equal to 10.32 (labeled as Sample I) and to 16.65 (labeled as Sample II) arc min, have been grown with velocities of $\sim$1.2 and $\sim$5.2 mm/h, respectively. Obtained single crystals exhibit (001) cleavage plane and the Sample I with better crystallographic quality has also well developed (100) natural planes.

The quantitative point analysis on the cleavage plane of the crystals was performed by Field Emission Scanning Electron Microscopy (FESEM) JEOL JSM 7600F operating at 20kV incident energy coupled with the Oxford INCA Energy Dispersive X-ray spectroscopy (EDX). Average chemical composition of the crystal matrix checked by Scanning Electron Microscopy (SEM) and EDX analysis (accuracy $\pm0.02$) is Fe$_{1.00}$Te$_{0.58}$Se$_{0.42}$ and Fe$_{1.01}$Te$_{0.57}$Se$_{0.43}$ for the Sample I and for the Sample II, respectively.

The magnetic measurements were carried out on single-crystalline samples of roughly rectangular shape, in the temperature range of 2$-$300 K, with magnetic field up to 50 kOe, using Quantum Design superconducting quantum interference device magnetometer. The magnetic field was applied parallel to the \textit{c}-axis of the crystal and to the \textit{ab} (001) plane, which is perpendicular to the \textit{c}-axis. Hydrostatic external pressure up to 11.3 kbar was applied, using easyLab Technologies Mcell 10 pressure cell with Daphne 7373 oil \cite{24}, being considered as the best pressure medium from the point of view of the smallest decrease of pressure with decreasing temperature, at least in the pressure range above 7 kbar \cite{25}. High-purity Sn wire (0.25 mm in diameter) was employed as an \textit{in situ} manometer. The background signal associated with the pressure cell was subtracted basing on the results obtained under ambient pressure for the sample placed in pressure cell and for the sample without pressure cell. We noted that the background contribution does not influence obtained results. The measurements of ac susceptibility (field amplitude 1 Oe, frequency 10 kHz) were performed with a Physical Property Measurement System (PPMS) of Quantum Design.

\section{Results and discussion}
\subsection{The critical temperature}

For single crystals of FeTe$_{0.5}$Se$_{0.5}$, noticeable differences between initial and estimated by EDX chemical composition as well as significant difference in FWHM of the 004 X-ray diffraction peak ($\Delta\omega$) are visible (see, for example, Ref. \cite{23}). Usually, they are attributed to a separation of phases with different Se/Te ratios, as reported in several papers \cite{9,26,27}. However, the data obtained for monophase single crystals of FeTe$_{0.65}$Se$_{0.35}$ \cite{23} indicated that the narrowest transition to the superconducting state (width $\sim$0.6 K) exhibit single crystals with relatively large values of $\Delta\omega$ equal to 6 arc min. Furthermore, the decrease in the $\Delta\omega$ value was found to be correlated with the increase of the width of the transition (90\%-10\% criterion). This correlation suggests that disorder in some sense enhances superconductivity in the FeTe$_{1-x}$Se$_{x}$ system, and properties of the superconducting state of FeTe$_{1-x}$Se$_{x}$ are very sensitive to the defects present in the sample \cite{23}.

\begin{figure}
\includegraphics{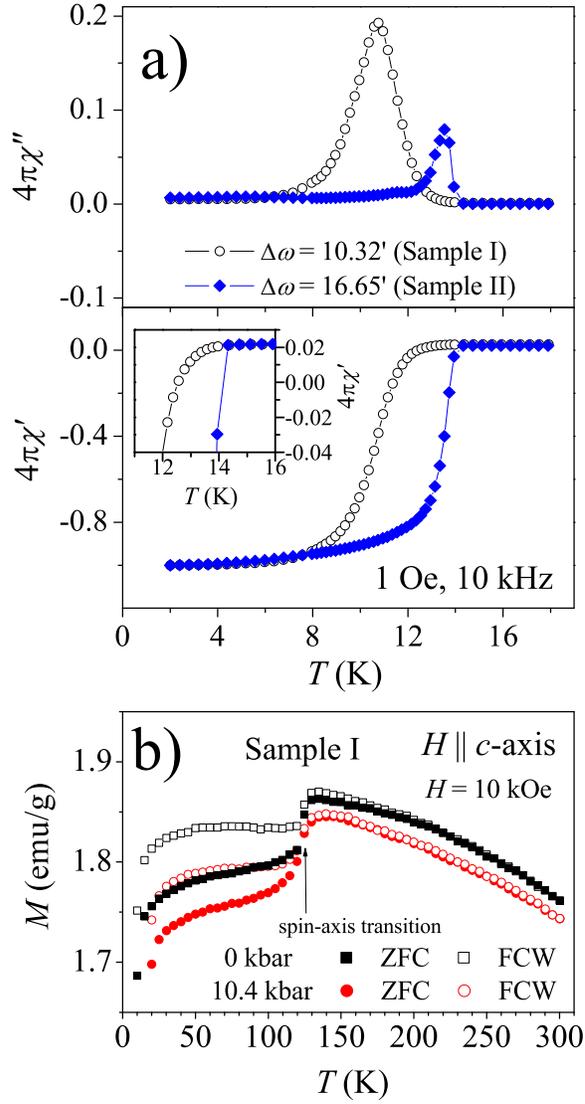}
\caption{(a) Temperature dependence of real part (lower panel) and imaginary part (upper panel) of ac magnetic susceptibility, normalized to the ideal value of -1 for the real part of ac susceptibility, measured in 1 Oe of ac field with 10 kHz in warming mode for two FeTe$_{0.5}$Se$_{0.5}$ single crystals of significantly different crystallographic quality, i.e., with different values of $\Delta\omega$ for 004 X-ray diffraction peak. Inset in the lower panel shows variation of $4\pi\chi'$ in the vicinity of ${T}_{\rm c}$. (b) Temperature dependence of dc magnetization measured in ZFC and FCW mode in magnetic field of 10 kOe, applied parallel to the \textit{c}-axis of the Sample I, at ambient pressure and under hydrostatic pressure of 10.4 kbar.}
\end{figure}

The main aim of our work was to study the pressure effect on intrinsic superconducting properties of FeTe$_{0.5}$Se$_{0.5}$. Since noticeable differences in the crystallographic quality were found among the crystals grown at various conditions, we decided to perform all of the measurements for two crystals of significantly different crystallographic quality, i.e., for the Sample I and for the Sample II with different $\Delta\omega$ value for 004 X-ray diffraction peak.

Figure 1a shows temperature dependence of real ($4\pi\chi'$) and imaginary ($4\pi\chi''$) parts of ac magnetic susceptibility measured in 1 Oe of ac field with 10 kHz in warming mode for two single crystals, grown with various cooling velocity and vertical gradient of temperature. Presented data were normalized to the ideal value of -1 for the real part of ac susceptibility for better comparison of the susceptibility data obtained for the samples with different shape and therefore subjected to different demagnetizing field. Inset in the lower panel of Fig. 1a shows variation of $4\pi\chi'$ for the crystals in the vicinity of ${T}_{\rm c}$. It is obvious that, despite of significant difference in the width of the transition to superconducting state, both of the crystals with different crystallographic quality are characterized by almost identical onset of ${T}_{\rm c}$.

Figure 1b presents temperature dependence of dc magnetization measured for the Sample I in wide temperature range up to 300 K in zero-field cooling (ZFC) and field-cooled warming (FCW) modes in magnetic field of 10 kOe, applied parallel to the \textit{c}-axis of the studied single crystal under ambient pressure and at hydrostatic pressure of 10.4 kbar. Similar behavior $-$ not shown $-$ was found for the Sample II. In the presented data, there is clearly visible transition at about 130 K, most likely related to spin reorientation of Fe$_{7}$Se$_{8}$-type minor phase or to Verwey transition in Fe$_{3}$O$_{4}$,\cite{28} coexisting in the crystal with the major tetragonal phase of FeTe$_{0.5}$Se$_{0.5}$ \cite{29,30}. The magnetization does not exceed 1.9 emu/g, therefore the volume fraction of Fe$_{7}$Se$_{8}$-type phase or of Fe$_{3}$O$_{4}$ should not be greater than few percent. Importantly, both temperature dependences of magnetization, at ambient and at hydrostatic pressure, are characterized by almost identical shape, indicating an absence of structural transition under pressure.

Temperature dependences of magnetic susceptibility in the vicinity of ${T}_{\rm c}$ for \textit{H} $\|$ \textit{c}-axis (upper panel) and for \textit{H} $\|$ \textit{ab}-plane (lower panel), measured under ambient pressure and applied hydrostatic pressure up to 10.4 kbar, in dc field of 10 Oe, for the Sample I, are presented in Fig. 2a. The critical temperature was defined as the point at \textit{x}-axis, where $M_{\rm ZFC}$($T$) curve deviates from constant, temperature independent background value. Almost linear dependence of ZFC magnetic susceptibility below ${T}_{\rm c}$, approximated well by parallel lines shifted to lower temperature with increasing pressure, indicates that superconducting transition width is almost unaffected by pressure, at least in the studied, relatively narrow, pressure range. A significant divergence between $M_{\rm ZFC}$ and $M_{\rm FCW}$ curves indicates relatively strong pinning of vortices for both \textit{H} $\|$ \textit{c}-axis and \textit{H} $\|$ \textit{ab}-plane even for the sample of better crystallographic quality (Fig. 2a). It was found that ${T}_{\rm c}$ increases linearly with pressure in the investigated pressure range from about 14 K at ambient pressure up to about 21 K at $P$ = 10.4 kbar, for both \textit{H} $\|$ \textit{c}-axis and \textit{H} $\|$ \textit{ab}-plane (upper panel of Fig. 2b). This confirms earlier reports on ${T}_{\rm c}$ increase, for FeTe$_{0.5}$Se$_{0.5}$ compound, in the pressure range of 0-10 kbar \cite{19,20,21}. The ${T}_{\rm c}$($P$) dependence for the Sample I, given in the upper panel of Fig. 2b by thick solid line with the pressure coefficient d${T}_{\rm c}$/d$P$ = 0.67(5) K/kbar, is a result of fitting of linear dependence with the least square method applied to the all of the data. Essentially identical data, within an experimental accuracy, were obtained for the Sample II, as it is presented in lower panel of Fig. 2b. The ${T}_{\rm c}$($P$) data for that sample are well approximated by linear dependence with the pressure coefficient d${T}_{\rm c}$/d$P$ = 0.69(2) K/kbar.

\begin{figure}
\includegraphics{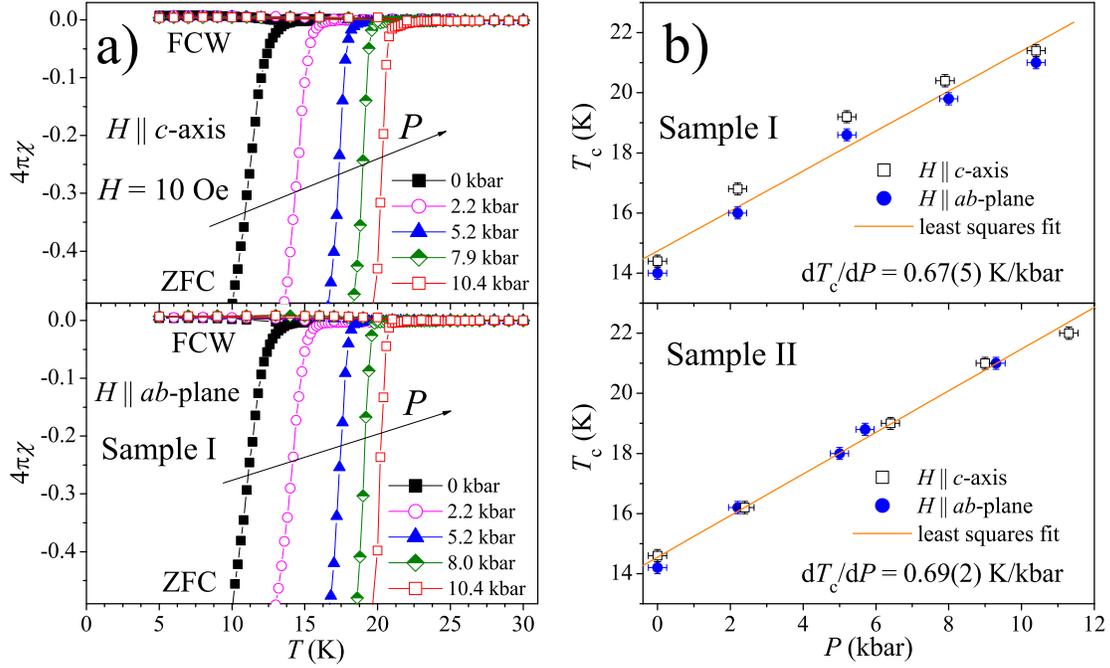}
\caption{(a) Temperature dependences of dc magnetic susceptibility in the vicinity of the critical temperature for \textit{H} $\|$ \textit{c}-axis (upper panel) and for \textit{H} $\|$ \textit{ab}-plane (lower panel), measured under ambient pressure and applied hydrostatic pressure up to 10.4 kbar, in dc field of 10 Oe, for the Sample I. (b) The pressure dependence of the critical temperature determined for both \textit{H} $\|$ \textit{c}-axis and \textit{H} $\|$ \textit{ab}-plane magnetic field configurations for the Sample I (upper panel) and for the Sample II (lower panel). The ${T}_{\rm c}$($P$) dependences given by solid lines are the results of fitting of linear dependence with the least square method.}
\end{figure}

\subsection{The thermodynamic parameters - the upper and the lower critical fields}

In order to estimate the change in the anisotropic thermodynamic parameters of the single crystal of  FeTe$_{0.5}$Se$_{0.5}$ subjected to hydrostatic pressure, we have evaluated temperature dependence of the upper and the lower critical fields in two geometries, \textit{H} $\|$ \textit{c}-axis and \textit{H} $\|$ \textit{ab}-plane (up to 50 kOe), for the studied samples under ambient pressure and at applied hydrostatic pressure of about 10 kbar.

Temperature dependence of magnetic moment \textit{m} measured under applied hydrostatic pressure of 10.4 kbar, for selected magnetic fields in the geometry \textit{H} $\|$ \textit{ab}-plane for the Sample I, is presented in Fig. 3a. From the above data we have determined ${T}_{\rm c2}$($H$=const), at the point where $m$($T$) deviates from linear temperature dependence, approximating well magnetic susceptibility in the normal state. The ${T}_{\rm c2}$($H$) data determined in this manner for various fields allowed us to plot ${H}_{\rm c2}$($T$) dependences for \textit{H} $\|$ \textit{c}-axis and for \textit{H} $\|$ \textit{ab}-plane for the studied samples under ambient pressure and at hydrostatic pressure of about 10 kbar.

\begin{figure}
\includegraphics{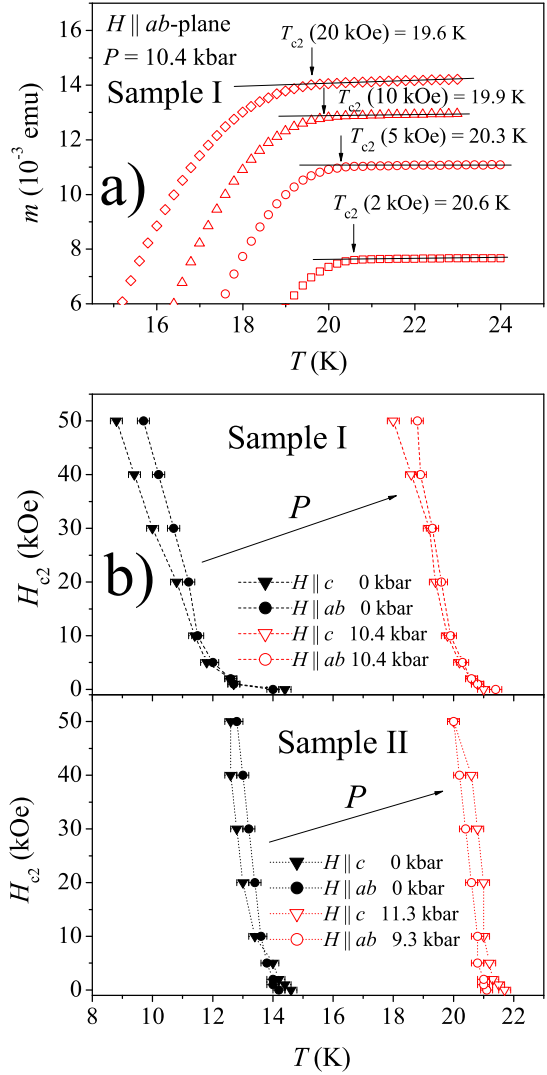}
\caption{(a) Temperature dependence of magnetic moment measured for the Sample I under applied hydrostatic pressure of 10.4 kbar, shown for selected magnetic fields for \textit{H} $\|$ \textit{ab}-plane. (b) Temperature dependences of the upper critical field for the Sample I for \textit{H} $\|$ \textit{c}-axis and \textit{H} $\|$ \textit{ab}-plane at ambient pressure and under hydrostatic pressure of 10.4 kbar (upper panel) and for the Sample II for \textit{H} $\|$ \textit{c}-axis and \textit{H} $\|$ \textit{ab}-plane at ambient pressure and under hydrostatic pressure of 11.3 kbar for \textit{H} $\|$ \textit{c}-axis and 9.3 kbar for \textit{H} $\|$ \textit{ab}-plane (lower panel).}
\end{figure}

Temperature dependences of the upper critical field for \textit{H} $\|$ \textit{c}-axis ($H^{\| c}_{\rm c2}$) and \textit{H} $\|$ \textit{ab}-plane ($H^{\| ab}_{\rm c2}$) for the Sample I at ambient pressure and under hydrostatic pressure of 10.4 kbar are shown in the upper panel of Fig. 3b. Significant increase of the upper critical field under pressure is clearly visible in this figure. Mainly, it results from the increase of ${T}_{\rm c}$ by about 7 K under pressure of 10.4 kbar. However, significant increase of the slope $-$d$H_{\rm c2}$/d$T$ in the linear part of $H_{\rm c2}$($T$) dependence is observed for higher fields. For lower fields, in the vicinity of ${T}_{\rm c}$, one can notice strong curvature. For \textit{H} $\|$ \textit{c}-axis, in the field range between 10 and 50 kOe, we have $-$d$H_{\rm c2}$/d$T$ = 15(1) kOe/K at ambient pressure, which rises up to 22(3) kOe/K under 10.4 kbar. In the case of \textit{H} $\|$ \textit{ab}-plane, an increase from 22(2) kOe/K ($P$ = 0 kbar) up to 34(3) kOe/K under pressure of 10.4 kbar is observed. The anisotropy of the slope $-$d$H_{\rm c2}$/d$T$ in the moderate fields, being equal to about 1.5 for the Sample I under ambient pressure and under pressure of 10.4 kbar, correlates quite well with the anisotropy of the penetration depth in the vicinity of ${T}_{\rm c}$ for single crystal of FeTe$_{0.5}$Se$_{0.5}$ investigated by Bendele \etal \cite{29} under ambient pressure. The estimation of zero-temperature value $H_{\rm c2}$(0) by extrapolation of the present data, covering a limited temperature range, down to low temperatures \cite{31} is not obvious because of strong curvature of $H_{\rm c2}$($T$) and possibly multi-band nature of the superconductivity. Nevertheless, assuming that the value of $H_{\rm c2}$(0) is proportional to ${T}_{\rm c}$ and to $-$d$H_{\rm c2}$/d$T$, determined in relatively wide field range above strong curvature of $H_{\rm c2}$($T$) in the vicinity of ${T}_{\rm c}$ \cite{31}, we can estimate a change of $H^{\| c}_{\rm c2}$(0) from 150 kOe under ambient pressure to 325 kOe under hydrostatic pressure of 10.4 kbar, what corresponds to a decrease of zero-temperature coherence length $\xi$$_{ab}$ from about 4.7 nm to 3.2 nm, according to relation \cite{32}:

\begin{equation}
H^{\| c}_{\rm c2} = \frac{\Phi_{0}}{2\pi\xi^{2}_{ab}},
\end{equation}
where $\Phi$$_{0}$ is elementary flux quantum and $\xi$$_{ab}$ is the coherence length in the \textit{ab}-plane.

Lower panel of Fig. 3b presents temperature dependences of the upper critical field for \textit{H} $\|$ \textit{c}-axis and \textit{H} $\|$ \textit{ab}-plane for the Sample II at ambient pressure and under hydrostatic pressure of about 10 kbar. Strong curvature of $H_{\rm c2}$($T$) in the vicinity of $T_{\rm c}$ noticed for the Sample I is much more suppressed for the sample with sharper transition to superconducting state (Sample II). Higher values of ${H}_{\rm c2}$ observed for \textit{H} $\|$ \textit{c}-axis under applied pressure of 11.3 kbar for this sample than those recorded for \textit{H} $\|$ \textit{ab}-plane under pressure of 9.3 kbar are due to the difference in the applied pressure and, therefore, due to the difference in ${T}_{\rm c}$ values. The slope $-$d$H_{\rm c2}$/d$T$, determined in the field range between 10 and 50 kOe, is much larger for the sample with sharp transition to superconducting state (Sample II). For the Sample II, we found the values of $-$d$H_{\rm c2}$/d$T$ equal to about 45(5) kOe/K for \textit{H} $\|$ \textit{c}-axis and to about 50(5) kOe/K for \textit{H} $\|$ \textit{ab}-plane, indicating much smaller value of the upper critical field anisotropy. Furthermore, the slope $-$d$H_{\rm c2}$/d$T$ is within experimental accuracy unchanged under pressure, suggesting that the increase of ${H}_{\rm c2}$ under pressure is directly related to the changes in ${T}_{\rm c}$ under pressure only. Presented data lead to an estimation of a change of $H^{\| c}_{\rm c2}$(0) from 450 kOe under ambient pressure to 690 kOe under hydrostatic pressure of 11.3 kbar, what corresponds to a decrease of zero-temperature $\xi$$_{ab}$ from about 2.7 nm to 2.2 nm. Relatively large values of ${H}_{\rm c2}$ and its small anisotropy for the Sample II most likely result from the extended amount of defects in the structure evidenced by wide X-ray peaks \cite{23} and therefore, they may not correspond to intrinsic ${H}_{\rm c2}$ values. On the other hand, the sample with larger amount of defects is characterized by stronger interband scattering and appearance of sufficiently strong interband scattering is an essential for enhanced superconducting state properties. Significant suppression of strong curvature of $H_{\rm c2}$($T$) in the vicinity of ${T}_{\rm c}$ for the sample with extended amount of defects may indicate the increasing interband scattering as a result of increasing structural inhomogeneity, consistently with observed increase of the upper critical field in the Sample II with extended inhomogeneity.

\begin{figure}
\includegraphics{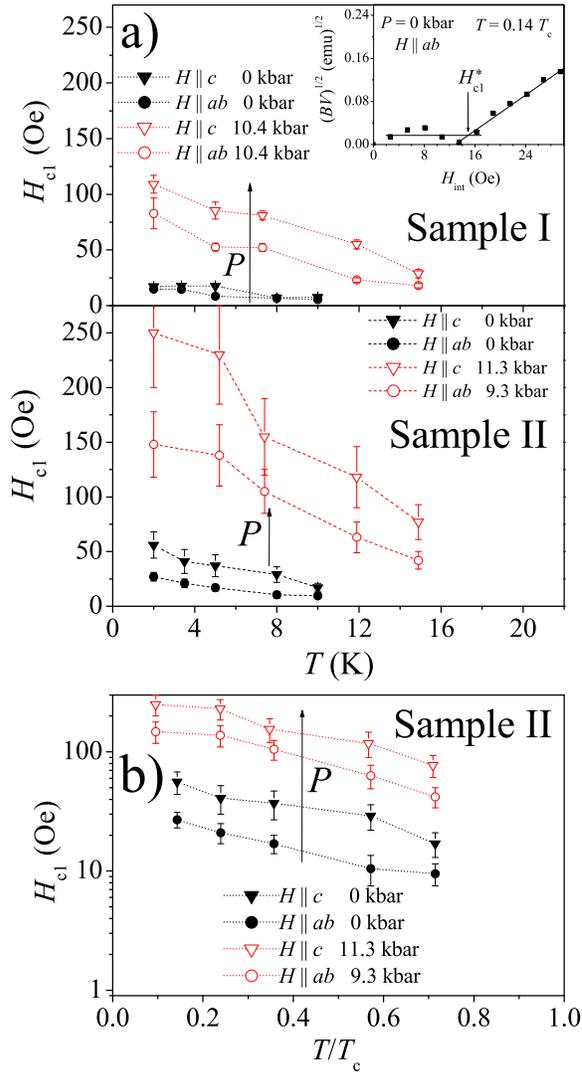}
\caption{(a) Temperature dependence of ${H}_{\rm c1}$ for the Sample I for \textit{H} $\|$ \textit{c}-axis and \textit{H} $\|$ \textit{ab}-plane determined at ambient pressure and under hydrostatic pressure of 10.4 kbar (upper panel) and for the Sample II at ambient pressure and under hydrostatic pressure of 11.3 kbar for \textit{H} $\|$ \textit{c}-axis and 9.3 kbar for \textit{H} $\|$ \textit{ab}-plane. Inset to the upper panel presents $(BV)^{1/2}$ vs. internal magnetic field, ${H}_{\rm int}$, determined at 2 K at ambient pressure for the Sample I for \textit{H} $\|$ \textit{ab}-plane. (b) The reduced temperature dependences of the lower critical field at ambient pressure and under hydrostatic pressure of 11.3 kbar for \textit{H} $\|$ \textit{c}-axis and 9.3 kbar for \textit{H} $\|$ \textit{ab}-plane, presented in semilogarithmic scale for the Sample II.}
\end{figure}

The temperature dependence of the lower critical field ${H}_{\rm c1}$ was studied by following the field $H^{*}_{\rm c1}$, for which the first vortices start to penetrate the sample at its surface, that is directly related to ${H}_{\rm c1}$ \cite{29}. The field dependences of the magnetic moment were measured at different temperatures for the magnetic field parallel to the \textit{ab}-plane and parallel to the \textit{c}-axis of the sample. For a given shape of the investigated crystal, the demagnetizing factors \textit{D} were calculated for the magnetic field applied along all of the crystallographic axis. The field $H^{*}_{\rm c1}$ was estimated according to the procedure introduced in Ref. \cite{33} and discussed in Ref. \cite{29}. The quantity $(BV)^{1/2}$ was calculated from the measured magnetic moment \textit{m} = \textit{MV} and plotted as a function of internal magnetic field ${H}_{\rm int}$ = ${H}_{\rm ext}$ $-$ $DM$, where ${H}_{\rm ext}$ denotes external magnetic field (see inset to the upper panel of Fig. 4a). Here, $B$ denotes the magnetic induction and \textit{V} is the sample volume. Since $B$ = $4\pi M$ $+$ ${H}_{\rm int}$ = $4\pi m/V$ + ${H}_{\rm int}$ = 0 in the Meissner state, it is possible to determine, from the data of $m(H_{\rm int})$, the field $H^{*}_{\rm c1}$ above which this equality is invalid. Hence, magnetic induction $B$ empirically scales as the square of $H$ above $H^{*}_{\rm c1}$, a plot of $(BV)^{1/2}$ as a function of ${H}_{\rm int}$  allows a straightforward determination of $H^{*}_{\rm c1}$. The sudden increase from zero occurs due to the penetration of vortices at $H^{*}_{\rm c1}$. For the case of weak bulk pinning, surface barrier may play a crucial role and determine the first field of flux penetration and the irreversibility line \cite{34,35,36}. The impact of surface barrier leads to asymmetric $M$($H$) loops. The descending branch is in such a case almost horizontal. For our samples, however, we observe symmetric magnetization loops, which means that bulk pinning controls mainly the entry and the exit of magnetic flux and therefore, we assume that $H^{*}_{\rm c1}$ is equal to $H_{\rm c1}$. Temperature dependence of ${H}_{\rm c1}$ for \textit{H} $\|$ \textit{c}-axis ($H^{\| c}_{\rm c1}$) and \textit{H} $\|$ \textit{ab}-plane ($H^{\| ab}_{\rm c1}$) for the Sample I determined at ambient pressure and under hydrostatic pressure of 10.4 kbar is presented in Fig. 4a. Data extrapolated to zero temperature are presented in Table I. Identical procedure was applied for the Sample II. Obtained data are presented in lower panel of Figure 4a. Obviously, the sample with narrow transition to superconducting state is characterized by larger values of ${H}_{\rm c1}$ (Sample II). It means that the penetration depth for this sample is smaller and the superconducting carrier density is bigger than that of the high crystallographic quality sample (Sample I).

Obtained data additionally indicate that structural disorder originating from kinetics of crystal growth process influences superconducting properties. In particular, our data support an observation that ions inhomogeneous spatial distribution enhances the superconductivity. Since the observed improvement of superconducting state properties is correlated with the suppression of a curvature of $H_{\rm c2}$($T$) in the vicinity of $T_{\rm c}$ one may suppose that an increase of interband scattering is directly responsible for the improvement of superconducting properties in the studied multiband superconductor.

\Table{\label{1}The pressure impact on the thermodynamic parameters describing superconducting state for both investigated single crystals of FeTe$_{0.5}$Se$_{0.5}$.}
\br
&\centre{2}{Sample I}&\centre{2}{Sample II} \\
&\crule{2}&\crule{2} \\
\ns
Quantity&0 kbar&10.4 kbar&0 kbar&11.3(*) or 9.3(**) kbar\\
\mr
${T}_{\rm c}$ (K) & 14.2(2) & 21.2(2) & 14.2(2) & 22.0(2)*  \\
-d$H^{\| c}_{\rm c2}$/d$T$ (kOe/K) & 15(1) & 22(3) & 45(5)  & 45(5)*   \\
-d$H^{\| ab}_{\rm c2}$/d$T$ (kOe/K) & 22(2) & 34(3) & 50(5) & 50(5)**   \\
 $H^{\| c}_{\rm c2}$(0) (kOe) & 150(10) & 325(45) & 450(50) & 690(80)* \\
 $H^{\| ab}_{\rm c2}$(0) (kOe) & 220(20) & 505(45) & 500(50) & 770(80)** \\
 $H^{\| c}_{\rm c1}$(0) (Oe) & 17(2) & 109(8) & 56(8)  & 250(30)*    \\
 $H^{\| ab}_{\rm c1}$(0) (Oe) & 15(2)& 83(14) & 27(5) & 150(30)**   \\
 $\xi$$_{ab}$(0) (nm) & 4.7(2) & 3.2(2) & 2.70(15) & 2.20(15)  \\
 $\xi$$_{c}$(0) (nm) & 3.9(2) & 2.55(15) & 2.55(15) & 2.05(15)  \\
 $\lambda$$_{ab}$(0) (nm) & 740(80) & 275(30) & 400(50) & 180(20)    \\
 $\lambda$$_{c}$(0) (nm) & 850(180) & 380(70) & 900(200) & 320(50)  \\
 $\kappa$$^{\| c}$(0) & 160(20) & 85(15) & 150(20)  & 80(15)    \\
 $\kappa$$^{\| ab}$(0) & 185(40) & 115(30) & 230(50) & 115(30)   \\
 \br
\end{tabular}
\end{indented}
\end{table}

From the data presented in lower panel of Fig. 4a, extrapolated
zero-temperature values for the Sample II, were found to be $H^{\|
ab}_{\rm c1}$(0) = 27(5) Oe and $H^{\| c}_{\rm c1}$(0) = 56(8) Oe at
ambient pressure and $H^{\| c}_{\rm c1}$(0) = 250(30) Oe under
pressure of 11.3 kbar while $H^{\| ab}_{\rm c1}$(0) = 150(30) Oe
under pressure of 9.3 kbar. The zero-temperature values of the lower
critical field for both field configurations correlate very well
with the values obtained by Bendele \etal for the single
crystals of identical nominal composition \cite{29}. The ${H}_{\rm
c1}$ increases significantly under applied external pressure for all
studied temperatures. The reduced temperature dependences of the
lower critical field at ambient pressure and under hydrostatic
pressure of 11.3 kbar for \textit{H} $\|$ \textit{c}-axis and 9.3
kbar for \textit{H} $\|$ \textit{ab}-plane, are presented in Fig. 4b
in semilogarithmic scale. The anisotropy of the lower critical field
($\gamma$$_{H \rm c1}$) does not increase under applied hydrostatic
pressure, the data presented in Fig. 4b rather indicate slight
decrease of $\gamma$$_{H \rm c1}$. In order to extract the values of
the magnetic penetration depth from the measured values of ${H}_{\rm
c1}$, the following basic relations were applied \cite{32}:

\begin{equation}
H^{\| c}_{\rm c1} = \frac{\Phi_{0}}{4\pi\lambda^{2}_{ab}}\left[\rm ln\left(\kappa^{\| c}\right) + 0.5\right],
\end{equation}

\begin{equation}
H^{\| ab}_{\rm c1} = \frac{\Phi_{0}}{4\pi\lambda_{ab}\lambda_{c}}\left[\rm ln\left(\kappa^{\| ab}\right) + 0.5\right].
\end{equation}
Here, $\lambda$$_{ab}$ and $\lambda$$_{c}$ denote the magnetic
penetration depths related to the superconducting current flowing in
the \textit{ab}-plane and along the \textit{c}-axis, respectively,
$\xi$$_{ab}$ and $\xi$$_{c}$ are the corresponding coherence
lengths, and $\kappa$$^{\| c}$ = $\lambda$$_{ab}$/$\xi$$_{ab}$ and
$\kappa$$^{\| ab}$ =
$(\lambda_{ab}\lambda_{c}/\xi_{ab}\xi_{c})^{1/2}$ are the
corresponding Ginzburg-Landau parameters. The zero temperature
values of $\xi$$_{ab}$(0) and $\xi$$_{c}$(0) at ambient pressure and
under hydrostatic pressure were derived from values of $H^{\|
c}_{\rm c2}$ and $H^{\| ab}_{\rm c2}$ extrapolated to zero
temperature for both field configurations. Then, for the Sample II,
the following zero-temperature values of the magnetic penetration
depths at ambient pressure were obtained: $\lambda$$_{ab}$(0)
$\approx$ 400(50) nm and $\lambda$$_{c}$(0) $\approx$ 900(200) nm.
These values are in a very good agreement with the values determined
by $\mu$SR measurements \cite{29}. The corresponding zero-temperature
values of the magnetic penetration depth at hydrostatic pressure of
about 10 kbar, are as follows: $\lambda$$_{ab}$(0) $\approx$ 180(20)
nm and $\lambda$$_{c}$(0) $\approx$ 320(50) nm. Obviously, estimated
low-temperature anisotropy of the penetration depth for
FeTe$_{0.5}$Se$_{0.5}$ under hydrostatic pressure is significantly
smaller than that one under ambient pressure. Furthermore, obtained
data suggest that anisotropy of $\lambda$ does not increase with
decreasing temperature, what is typical for chalcogenides at ambient
pressure. However, obtained data are insufficient to make conclusive
statement concerning temperature dependence of the anisotropy of the
penetration depth in FeTe$_{0.5}$Se$_{0.5}$ under pressure. Summary
of the changes of thermodynamic parameters under pressure for both
studied samples is given in Table I.

\begin{figure}
\includegraphics{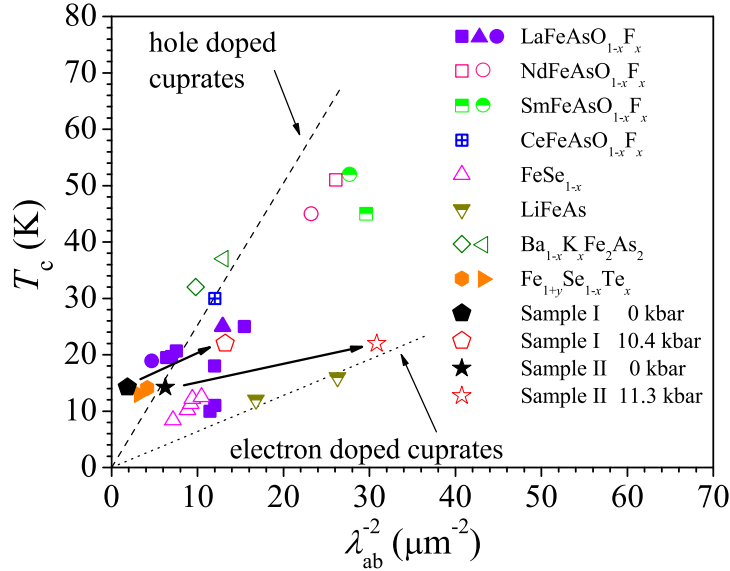}
\caption{Pressure impact on the position of FeTe$_{0.5}$Se$_{0.5}$ at the Uemura plot of a selection of some Fe-based high-temperature superconductors (after Ref. \cite{29}). The arrows indicate change of the position in the plot of the investigated crystals of significantly different crystallographic quality when subjected to hydrostatic pressure of 10.4 kbar (Sample I) and 11.3 kbar (Sample II). The Uemura relation observed for underdoped cuprates is included for comparison as a dashed line for hole doping and as a dotted line for electron doping.}
\end{figure}

Uemura \etal \cite{37} have found an empirical relation between the zero-temperature superconducting carrier density ${\rho}_{\rm s}$(0) $\propto$ $\lambda$$^{-2}_{ab}$(0) and ${T}_{\rm c}$ which seems to be generic for various families of cuprate high-temperature superconductors (Uemura plot). This "universal" relation ${T}_{\rm c}$(${\rho}_{\rm s}$) has the following features: with increasing carrier doping ${T}_{\rm c}$ initially increases linearly [${T}_{\rm c}$ $\propto$ ${\rho}_{\rm s}$(0)], then saturates, and finally is suppressed for high carrier doping. It is interesting to check, how the Uemura relation holds for iron-based superconductors subjected to hydrostatic pressure. For this reason, ${T}_{\rm c}$ vs. $\lambda$$^{-2}_{ab}$(0) is plotted in Fig. 5 for a selection of various Fe-based superconductors investigated so far, \cite{29,38,39,40,41,42,43,44,45,46,47,48} together with the pressure impact on the position of both FeTe$_{0.5}$Se$_{0.5}$ samples investigated in this work. The figure was prepared using the values of ${T}_{\rm c}$($\lambda$$_{ab}$(0)) obtained for the Sample I at ambient pressure and under hydrostatic pressure of 10.4 kbar and for the Sample II at ambient pressure and under hydrostatic pressure of 11.3 kbar. The Uemura relation observed for underdoped cuprates is included for comparison as a dashed line for hole doping and as a dotted line for electron doping. The penetration depth values obtained under ambient pressure locate the studied samples in the area of hole-doped compounds. An application of hydrostatic pressure of about 10 kbar shifts the position of studied samples in the diagram ${T}_{\rm c}$($\lambda$$_{ab}$(0)) towards the area of electron-doped compounds, instead of the shift along the line denoting hole-doped compounds. The effect is very well visible for the Sample II placed almost ideally on the line denoting hole-doped compounds at ambient pressure as well as on the line denoting electron-doped compounds under hydrostatic pressure of 11.3 kbar. The Sample I, despite of essentially identical value of ${T}_{\rm c}$ as the Sample II, is characterized by much higher value of $\lambda$$_{ab}$(0) both at ambient and under hydrostatic pressure, and therefore its position in the Uemura plot is shifted towards the lower $\lambda$$_{ab}^{-2}$ values as compared to those expected for hole-doped and electron-doped compounds, respectively. Obviously, for both studied samples the external pressure affects the density of superconducting carriers. However, it may cause also an induction of magnetic phase, similar to that reported by Bendele \etal \cite{49} in FeSe crystal, manifested by ${T}_{\rm c}$($P$) dependence not going along the hole-doped compounds line. Importantly, we noticed more pronounced increase in superconducting carrier density under pressure than that in  the critical temperature, what may indicate an appearance of a mechanism limiting the increase of ${T}_{\rm c}$ with pressure. However, we should note that the change of lattice constants under pressure leads to the change of superconducting carrier effective mass what affects values of $\lambda$$_{ab}$(0).

\begin{figure}
\includegraphics{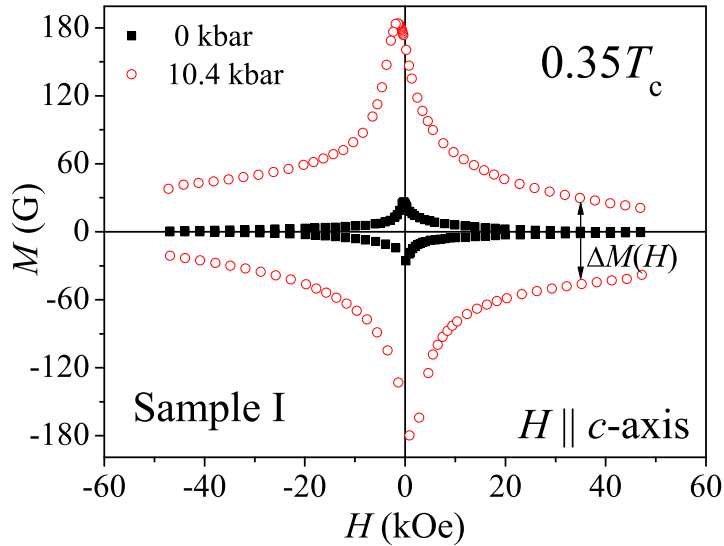}
\caption{Hysteresis loop of single crystal of FeTe$_{0.5}$Se$_{0.5}$ for \textit{H} $\|$ \textit{c}-axis recorded at 5 K and at 7.3 K for the Sample I at ambient pressure and under hydrostatic pressure of 10.4 kbar, respectively, i.e., at the same reduced temperature of 0.35${T}_{\rm c}$.}
\end{figure}

\begin{figure*}
\includegraphics{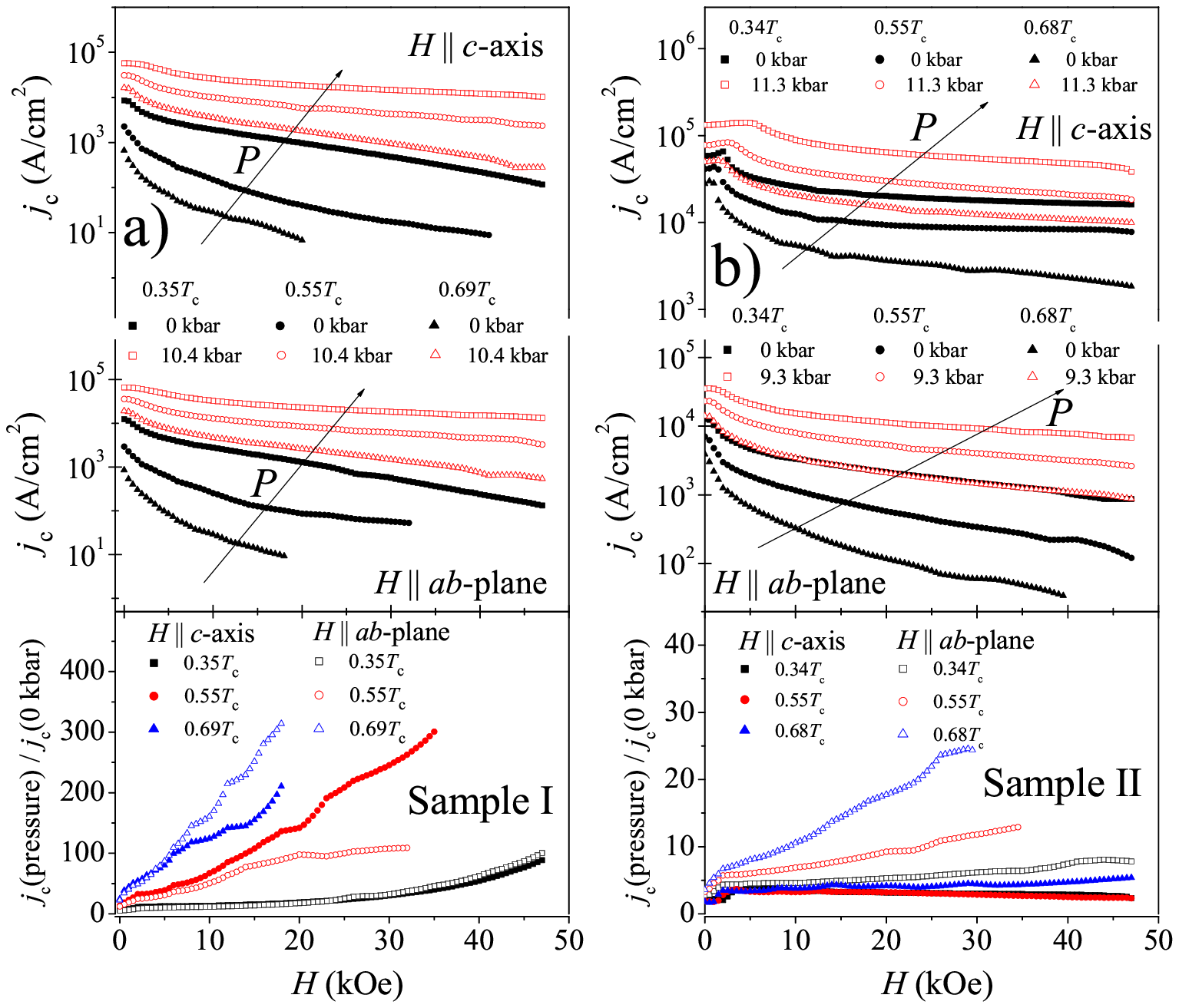}
\caption{(a) Magnetic field dependence of the critical current density in semilogarithmic scale at ambient pressure and under hydrostatic pressure of 10.4 kbar for the Sample I, at various temperatures for \textit{H} $\|$ \textit{c}-axis (upper panel) and  \textit{H} $\|$ \textit{ab}-plane (middle panel). Lower panel: Field dependence of the enhancement of the critical current density under pressure, i.e., of the ratio of the critical current densities under hydrostatic pressure of 10.4 kbar and at ambient pressure for the Sample I at reduced temperatures of 0.35, 0.55, and 0.69${T}_{\rm c}$ in magnetic field \textit{H} $\|$ \textit{c}-axis and \textit{H} $\|$ \textit{ab}-plane. (b) The same as in Fig. 7a  for the Sample II at ambient pressure and under hydrostatic pressure of 11.3 kbar for \textit{H} $\|$ \textit{c}-axis and 9.3 kbar for \textit{H} $\|$ \textit{ab}-plane at reduced temperatures of 0.34, 0.55, and 0.68${T}_{\rm c}$.}
\end{figure*}

\subsection{The critical current density}

Hysteresis loops of the studied single crystals were recorded at various temperatures in magnetic field applied along both \textit{H} $\|$ \textit{c}-axis and \textit{H} $\|$ \textit{ab}-plane at ambient pressure and under hydrostatic pressure of about 10 kbar. Figure 6 presents typical data recorded for the Sample I for \textit{H} $\|$ \textit{c}-axis at 5 K at ambient pressure and at 7.3 K under hydrostatic pressure of 10.4 kbar, i.e., at the same reduced temperature of 0.35${T}_{\rm c}$. Using Bean's model \cite{50,51}, for the sample of rectangular shape, one can estimate the superconducting critical current density according to the formula:

\begin{equation}
j_{\rm c}(H) = \frac{20\Delta\textit{M}(H)}{a\left(1-\frac{a}{3b}\right)}.
\end{equation}
Here, $\Delta$\textit{M} (in Gauss) is the width of the hysteresis loop (see, Fig. 6), \textit{a} and \textit{b} are the sample dimensions (in cm) in the plane perpendicular to applied magnetic field and the critical current density is in A/cm$^{2}$. Magnetic field dependence of the critical current density for the Sample I at ambient pressure and under hydrostatic pressure of 10.4 kbar, calculated according to the Eq. (4), for all of the studied temperatures in magnetic field geometry \textit{H} $\|$ \textit{c}-axis and \textit{H} $\|$ \textit{ab}-plane is presented in Fig. 7a (upper and middle panels). We note relatively small value of the estimated critical current density, $j_{\rm c}$,  as compared to those observed in single-crystalline iron pnictides \cite{52}. However, the obtained $j_{\rm c}$ values are not surprising since it was shown that FeTe$_{0.5}$Se$_{0.5}$ may exhibit the coexistence of two tetragonal phases \cite{9,26,27}. The presence of such phases lowers the transport current density as phase separation boundaries prevent to develop a global circulating current \cite{29}. This leads to a relatively low value of magnetic critical current density, when calculated taking into account the diameter of the sample. Furthermore, both, the upper and the lower, critical fields for the Sample I are quite small in comparison with those for the Sample II (see, Figs. 3b and 4a and Table I) and pinning is expected to be proportional to the thermodynamic critical field. Field dependence of the increase of the critical current density under pressure, i.e., of the ratio of critical current densities under hydrostatic pressure of 10.4 kbar and at ambient pressure at reduced temperatures of 0.35, 0.55, and 0.69${T}_{\rm c}$ in magnetic field \textit{H} $\|$ \textit{c}-axis and \textit{H} $\|$ \textit{ab}-plane is presented in lower panel of Fig. 7a. The critical current density strongly increases under pressure by at least one order of magnitude, for \textit{H} $\|$ \textit{c}-axis and \textit{H} $\|$ \textit{ab}-plane, for all investigated reduced temperatures and in full magnetic field range (lower panel of Fig. 7a). It can be explained by an improvement of the effectiveness of small defects in the sample subjected to pressure, because of a decrease of the coherence length under pressure, and by an increase of the thermodynamic critical field under pressure due to the increase of both the lower and the upper critical fields. The influence of pressure on $j_{\rm c}$ is evidently stronger at higher magnetic fields, up to two orders of magnitude (lower panel of Fig. 7a). It is not surprising since significant increase of ${H}_{\rm c2}$ under pressure was noted too.

Magnetic field dependence of the critical current density for the Sample II at ambient pressure and under hydrostatic pressure of 11.3 kbar in magnetic field geometry \textit{H} $\|$ \textit{c}-axis and of 9.3 kbar in the \textit{H} $\|$ \textit{ab}-plane geometry for all of the studied temperatures is presented in Fig. 7b (upper and middle panels). Field dependence of the increase of the critical current density under pressure is presented in lower panel of Fig. 7b. The Sample II is characterized by significantly enhanced critical current density at ambient pressure, as compared to the Sample I, because of extended amount of defects in the structure, evidenced by relatively wide X-ray peaks. Consequently, the increase of the critical current density under pressure is strongly reduced for the Sample II, especially in the geometry \textit{H} $\|$ \textit{c}-axis and at low temperatures, where the initial critical current density is the highest.

\section{Conclusions}

The magnetic studies at ambient and under hydrostatic pressure were performed for single crystals of FeTe$_{0.5}$Se$_{0.5}$ in order to investigate the pressure impact on basic parameters of the superconducting state. We compared influence of hydrostatic pressure on the properties of two crystals with significantly different amount of defects. We have found pressure-induced enhancement of all investigated parameters. Furthermore, we noted that the application of hydrostatic pressure does not increase the anisotropy of superconducting state parameters. However, more  pronounced increase in superconducting carrier density under pressure than that in critical temperature was found, indicating an appearance of a mechanism limiting the increase of ${T}_{\rm c}$ with pressure.

Comparison of pressure impact on superconducting properties of two samples with different amount of defects leads to the following conclusion: significant suppression of strong curvature of $H_{\rm c2}$($T$) in the vicinity of ${T}_{\rm c}$ for the sample with extended amount of defects indicates the increasing interband scattering as a result of increasing structural inhomogeneity. Since the suppression of the curvature of $H_{\rm c2}$($T$) in the vicinity of $T_{\rm c}$ is correlated with observed improvement of superconducting state properties one may suppose that an increase of interband scattering is directly responsible for the improvement of superconducting properties in the studied multiband superconductor. It may explain the origin of relatively poor superconducting state properties of the single crystals of better crystallographic quality.

\ack

This work was supported by the EC through the FunDMS Advanced Grant of the European Research Council (FP7 'Ideas') and by National Science Centre (Poland) based on decision No. DEC-2011/01/B/ST3/02374. We thank Tomasz Dietl for suggesting this research and valuable discussions.


\Bibliography{50}

\bibitem{1} Kamihara Y, Watanabe T, Hirano M and Hosono H 2008 \textit{J. Am. Chem. Soc.} \textbf{130} 3296
\bibitem{2} Rotter M, Tegel M and Johrendt D 2008 \textit{Phys. Rev. Lett.} \textbf{101} 107006
\bibitem{3} Wang X C, Liu Q Q, Lv Y X, Gao W B, Yang L X, Yu R C, Li F Y and Jin C Q 2008 \textit{Solid State Commun.} \textbf{148} 538
\bibitem{4} Hsu F-C, Luo J-Y, Yeh K-W, Chen T-K, Huang T-W, Wu P M, Lee Y-C, Huang Y-L, Chu Y-Y, Yan D-C and Wu M-K 2008 \textit{Proc. Natl. Acad. Sci. USA} \textbf{105} 14262
\bibitem{5} Yeh K-W, Huang T-W, Huang Y-L, Chen T-K, Hsu F-C, Wu P M, Lee Y-C, Chu Y-Y, Chen C-L, Luo J-Y, Yan D-C and Wu M-K 2008 \textit{Europhys. Lett.} \textbf{84} 37002
\bibitem{6} Ogino H, Matsumura Y, Katsura Y, Ushiyama K, Horii S, Kishio K and Shimoyama J 2009 \textit{Supercond. Sci. Technol.} \textbf{22} 075008 
\bibitem{7} Fang M H, Pham H M, Qian B, Liu T J, Vehstedt E K, Liu Y, Spinu L and Mao Z Q 2008 \textit{Phys. Rev. B} \textbf{78} 224503
\bibitem{8} Bao W, Qiu Y, Huang Q, Green M A, Zajdel P, Fitzsimmons M R, Zhernenkov M, Chang S, Fang M, Qian B, Vehstedt E K, Yang J, Pham H M, Spinu L and Mao Z Q 2009 \textit{Phys. Rev. Lett.} \textbf{102} 247001
\bibitem{9} Khasanov R, Bendele M, Amato A, Babkevich P, Boothroyd A T, Cervellino A, Conder K, Gvasaliya S N, Keller H, Klauss H-H, Luetkens H, Pomjakushin V, Pomjakushina E and Roessli B 2009 \textit{Phys. Rev. B} \textbf{80} 140511
\bibitem{10} Mizuguchi Y, Tomioka F, Tsuda S, Yamaguchi T and Takano Y 2008 \textit{Appl. Phys. Lett.} \textbf{93} 152505
\bibitem{11} Medvedev S, McQueen T M, Troyan I A, Palasyuk T, Eremets M I, Cava R J, Naghavi S, Casper F, Ksenofontov V, Wortmann G and Felser C 2009 \textit{Nature Mater.} \textbf{8} 630
\bibitem{12} Garbarino G, Sow A, Lejay P, Sulpice A, Toulemonde P, Mezouar M and Nunez-Regueiro M 2009 \textit{Europhys. Lett.} \textbf{86} 27001
\bibitem{13} Braithwaite D, Salce B, Lapertot G, Bourdarot F, Marin C, Aoki D and Hanfland M 2009 \textit{J. Phys.: Condens. Matter} \textbf{21} 232202
\bibitem{14} Guo J-G, Jin S-F, Wang G, Wang S-C, Zhu K-X, Zhou T-T, He M and Chen X-L 2010 \textit{Phys. Rev. B} \textbf{82} 180520
\bibitem{15} Krzton-Maziopa A, Shermadini Z, Pomjakushina E, Pomjakushin V, Bendele M, Amato A, Khasanov R, Luetkens H and Conder K 2011 \textit{J. Phys.: Condens. Matter} \textbf{23} 052203
\bibitem{16} Li C-H, Shen B, Han F, Zhu X and Wen H-H 2011 \textit{Phys. Rev. B} \textbf{83} 184521
\bibitem{17} Guo J, Chen X, Zhang C, Guo J, Chen X, Wu Q, Gu D, Gao P, Dai X, Yang L, Mao H-K, Sun L and Zhao Z 2011 Pressure-driven quantum criticality in an iron-selenide superconductor \textit{Preprint} cond-mat.supr-con/1101.0092
\bibitem{18} Seyfarth G, Jaccard D, Pedrazzini P, Krzton-Maziopa A, Pomjakushina E, Conder K and Shermadini Z 2011 \textit{Solid State Commun.} \textbf{151} 747
\bibitem{19} Horigane K, Takeshita N, Lee Ch-H, Hiraka H and Yamada K 2009 \textit{J. Phys. Soc. Jpn.} \textbf{78} 063705
\bibitem{20} Tsoi G, Stemshorn A K, Vohra Y K, Wu P M, Hsu F C, Huang Y L, Wu M K, Yeh K W and Weir S T 2009 \textit{J. Phys.: Condens. Matter} \textbf{21} 232201
\bibitem{21} Huang Ch-L, Chou Ch-Ch, Tseng K-F, Huang Y-L, Hsu F-Ch, Yeh K-W, Wu M-K and Yang H-D 2009 \textit{J. Phys. Soc. Jpn.} \textbf{78} 084710
\bibitem{22} Stemshorn A K, Vohra Y K, Wu P M, Hsu F C, Huang Y L, Wu M K and Yeh K W 2009 \textit{High Pressure Res.} \textbf{29} 267
\bibitem{23} Gawryluk D J, Fink-Finowicki J, Wisniewski A, Puzniak R, Domukhovski V, Diduszko R, Kozlowski M and Berkowski M 2011 \textit{Supercond. Sci. Technol.} \textbf{24} 065011
\bibitem{24} Murata K, Yoshino H, Yadav H O, Honda Y and Shirakava N 1997 \textit{Rev. Sci. Instrum.} \textbf{68} 2490
\bibitem{25} Kamar\'{a}d J, Mach\'{a}tov\'{a} A and Arnold Z 2004 \textit{Rev. Sci. Instrum.} \textbf{75} 5022
\bibitem{26} Sales B C, Sefat A S, McGuire M A, Jin R Y, Mandrus D and Mozharivskyj Y 2009 \textit{Phys. Rev. B} \textbf{79} 094521
\bibitem{27} Lumsden M D, Christianson A D, Goremychkin E A, Nagler S E, Mook H A, Stone M B, Abernathy D L, Guidi T, MacDougall G J, de la Cruz C, Sefat A S, McGuire M A, Sales B C and Mandrus D 2010 \textit{Nature Phys.} \textbf{6} 182
\bibitem{28} Verwey E J W 1939 \textit{Nature} \textbf{144} 327
\bibitem{29} Bendele M, Weyeneth S, Puzniak R, Maisuradze A, Pomjakushina E, Conder K, Pomjakushin V, Luetkens H, Katrych S, Wisniewski A, Khasanov R and Keller H 2010 \textit{Phys. Rev. B} \textbf{81} 224520
\bibitem{30} Szymanski K, Olszewski W, Dobrzynski L, Satula D, Gawryluk D J, Berkowski M, Puzniak R and Wisniewski A 2011 \textit{Supercond. Sci. Techn.} \textbf{24} 105010
\bibitem{31} Werthamer N R, Helfand E and Hohenberg P C 1966 \textit{Phys. Rev.} \textbf{147} 295
\bibitem{32} Tinkham M 1975 \textit{Introduction to Superconductivity} (Krieger, Malabar, Florida)
\bibitem{33} Naito M, Matsuda A, Kitazawa K, Kambe S, Tanaka I and Kojima H 1990 \textit{Phys. Rev. B} \textbf{41} 4823
\bibitem{34} Clem J R 1974 \textit{Proceeding of the 13th Conference on Low Temperature Physics (LT 13)} vol. 1 ed K D Timmerhaus, W J O'Sullivan and E F Hammel (New York: Plenum) p~102
\bibitem{35} Burlachkov L 1993 \textit{Phys. Rev. B} \textbf{47} 8056
\bibitem{36} Burlachkov L, Geshkenbein V B, Koshelev A E, Larkin A I and Vinokur V M 1994 \textit{Phys. Rev. B} \textbf{50} 16770
\bibitem{37} Uemura Y J, Luke G M, Sternlieb B J, Brewer J H, Carolan J F, Hardy W N, Kadono R, Kempton J R, Kiefl R F, Kreitzman S R, Mulhern P, Riseman T M, Williams D L, Yang B X, Uchida S, Takagi H, Gopalakrishnan J, Sleight A W, Subramanian M A, Chien C L, Cieplak M Z, Xiao G, Lee V Y, Statt B W, Stronach C E, Kossler W J and Yu X H 1989 \textit{Phys. Rev. Lett.} \textbf{62} 2317 
\bibitem{38} Luetkens H, Klauss H-H, Kraken M, Litterst F J, Dellmann T, Klingeler R, Hess C, Khasanov R, Amato A, Baines C, Kosmala M, Schumann O J, Braden M, Hamann-Borrero J, Leps N, Kondrat A,  Behr G, Werner J and B\"{u}chner B 2009 \textit{Nature Mater.} \textbf{8} 305
\bibitem{39} Drew A J, Niedermayer Ch, Baker P J, Pratt F L, Blundell S J, Lancaster T, Liu R H, Wu G, Chen X H, Watanabe I, Malik V K, Dubroka A, R\"{o}ssle M, Kim K W, Baines C and Bernhard C 2009 \textit{Nature Mater.} \textbf{8} 310
\bibitem{40} Khasanov R, Conder K, Pomjakushina E, Amato A, Baines C, Bukowski Z, Karpinski J, Katrych S, Klauss H-H, Luetkens H, Shengelaya A and Zhigadlo N D 2008 \textit{Phys. Rev. B} \textbf{78} 220510
\bibitem{41} Kim H, Martin C, Gordon R T, Tanatar M A, Hu J, Qian B, Mao Z Q, Hu R, Petrovic C, Salovich N, Giannetta R and Prozorov R 2010 \textit{Phys. Rev. B} \textbf{81} 180503(R)
\bibitem{42} Luetkens H, Klauss H-H, Khasanov R, Amato A, Klingeler R, Hellmann I, Leps N, Kondrat A, Hess C, K\"{o}hler A, Behr G, Werner J and B\"{u}chner B 2008 \textit{Phys. Rev. Lett.} \textbf{101} 097009
\bibitem{43} Takeshita S and Kadono R 2009 \textit{New J. Phys.} \textbf{11} 035006
\bibitem{44} Carlo J P, Uemura Y J, Goko T, MacDougall G J, Rodriguez J A, Yu W, Luke G M, Dai P, Shannon N, Miyasaka S, Suzuki S, Tajima S, Chen G F, Hu W Z, Luo J L and Wang N L 2009 \textit{Phys. Rev. Lett.} \textbf{102} 087001
\bibitem{45} Khasanov R, Luetkens H, Amato A, Klauss H-H, Ren Z-A, Yang J, Lu W and Zhao Z-X 2008 \textit{Phys. Rev. B} \textbf{78} 092506
\bibitem{46} Khasanov R, Bendele M, Amato A, Conder K, Keller H, Klauss H-H, Luetkens H and Pomjakushina E 2010 \textit{Phys. Rev. Lett.} \textbf{104} 087004
\bibitem{47} Pratt F L, Baker P J, Blundell S J, Lancaster T, Lewtas H J, Adamson P, Pitcher M J, Parker D R and Clarke S J 2009 \textit{Phys. Rev. B} \textbf{79} 052508
\bibitem{48} Khasanov R, Evtushinsky D V, Amato A, Klauss H-H, Luetkens H, Niedermayer Ch, B\"{u}chner B, Sun G L, Lin C T, Park J T, Inosov D S and Hinkov V 2009 \textit{Phys. Rev. Lett.} \textbf{102} 187005
\bibitem{49} Bendele M, Amato A, Conder K, Elender M, Keller H, Klauss H-H, Luetkens H, Pomjakushina E, Raselli A and Khasanov R 2010 \textit{Phys. Rev. Lett.} \textbf{104} 087003
\bibitem{50} Bean C P 1962 \textit{Phys. Rev. Lett.} \textbf{8} 250
\bibitem{51} Bean C P 1964 \textit{Rev. Mod. Phys.} \textbf{36} 31
\bibitem{52} Karpinski J, Zhigadlo N D, Katrych S, Bukowski Z, Moll P, Weyeneth S, Keller H, Puzniak R, Tortello M, Daghero D, Gonnelli R, Maggio-Aprile I, Fasano Y, Fischer O, Rogacki K and Batlogg B 2009 \textit{Physica C} \textbf{469} 370

\endbib
\end{document}